\documentclass[11pt]{article}
\usepackage{moriond,epsfig}

\def\MyPhoto{\pdfimage width 35mm {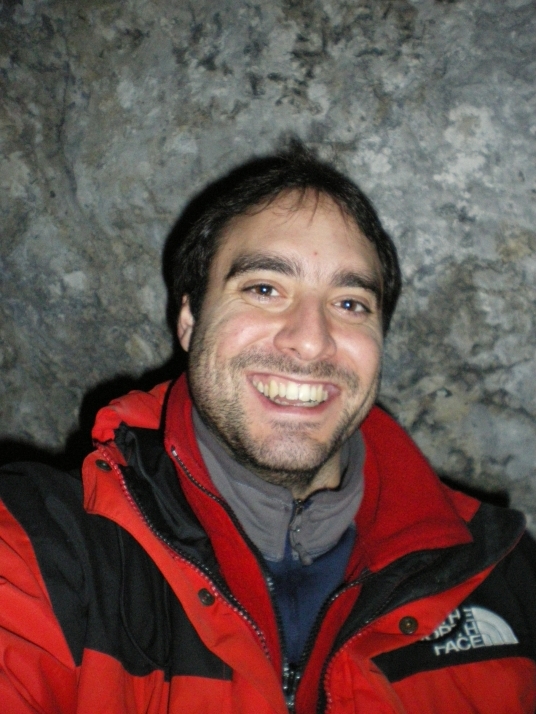}}

\def\Journal#1#2#3#4{{#1} {\bf #2}, #3 (#4)}

\def\NPB{{\em Nucl. Phys.} B}
\def\PLB{{\em Phys. Lett.} B}
\def\PRL{\em Phys. Rev. Lett.}
\def\APP{\em Astropart. Phys.}
\def\PRD{{\em Phys. Rev.} D}
\def\PRC{{\em Phys. Rev.} C}

\def\SSR{\em Space Sci. Rev.}
\def\AJS{\em Astrophys. J. Suppl.}
\def\SNP{\em Sov. J. Nucl. Phys.}
\def\JCAP{\em JCAP}
\def\JHEP{\em Jour. High En. Phys.}
\def\JPG{{\em Jour. Phys.} G}

\def\be{\begin{equation}}
\def\ee{\end{equation}}
\def\bea{\begin{eqnarray}}
\def\eea{\end{eqnarray}}

\begin{document}
\vspace*{4cm}
\title{RESULTS FROM THE BOREXINO EXPERIMENT}

\author{TIMO LEWKE \\ on behalf of the Borexino Collaboration}

\address{Physik Department, Technische Universit\"at M\"unchen \\
85747 Garching, Germany}

\maketitle\abstracts{
Borexino is a low threshold liquid-scintillator detector for solar neutrinos located in the LNGS underground laboratory, Italy.
Because of the ultra-high radio purity it is the first experiment able to do a real-time analysis of the low energetic solar neutrinos.
A detection of the solar $^{7}$Be neutrinos with a rate of $49\pm7$ counts/day/100tons can be reported (192 days of live time measurement).
$^{8}$B neutrinos are observed with a preliminary rate of $0.26\pm0.06$ counts/day/100tons after 246 live days.
All detected neutrino fluxes agree with the SSM predictions in case of the MSW-LMA oscillation solution. Borexino is the first experiment
with the ablility to simultaneously measure solar neutrino oscillation in the vacuum-dominated and the matter-enhanced energy regions.}

\section{Introduction}\label{sec:intro}
In the past years, neutrinos have been studied very intensely. Neutrino oscillations were observed, for example via SNO~\cite{aha} and Super-Kamiokande~\cite{hos},
and the solar neutrino oscillation parameters $\Delta m^{2}_{21}=7.59^{+0.21}_{-0.21} \cdot 10^{-5}eV^{2}$
and $\tan^{2}{\Theta_{12}}=0.47^{+0.06}_{-0.05}$~\cite{abe} could be determined. Nervertheless no experiment of the so far used water-Cherenkov experiments was able
to measure neutrino energies lower than 5 MeV due to their threshold.\newline
With Borexino as an organic liquid scintillator detector, a new field of low energetic neutrino physics can be entered. Because of the low threshold, $^{7}$Be neutrinos can be
observed in real-time measurement~\cite{arp1,arp2}.\newline
The results reported in this work present the real-time measurement of $^{7}$Be and $^{8}$B solar neutrinos in Borexino.
Therefore the first simultaneous measurement of the neutrino survival probability $P_{ee}$ out of both, vacuum and matter enhanced oscillation regions, can be announced.
After a short description of the detector layout and the radioactive purity levels achieved in the experiment, this work will present the
results of the measurements in Borexino. It will conclude with a brief outlook.

\section{The Borexino Detector}\label{sec:deetector}
\begin{figure}[h]
\begin{center}
\begin{minipage}{0.8\textwidth}
\centering{\includegraphics[width=0.75\textwidth]{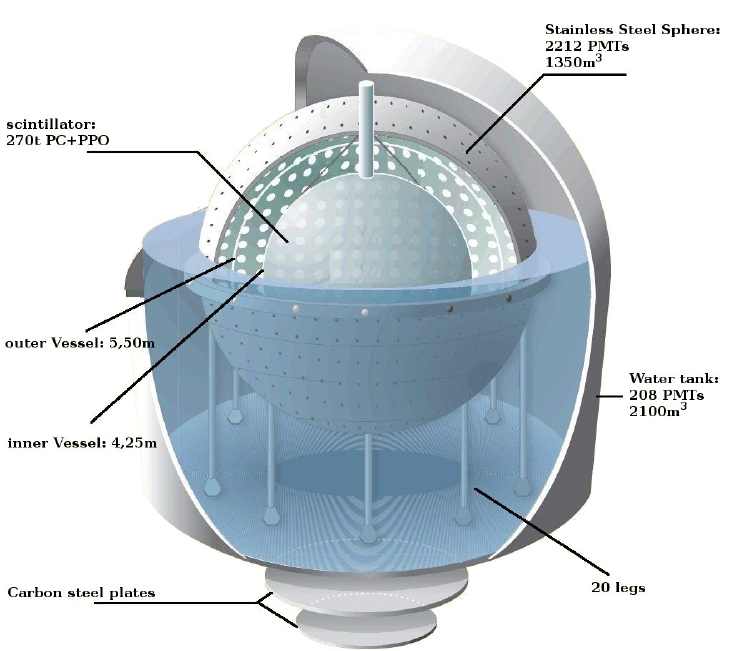}}
\caption{Shematic view of the Borexino detector.}\label{shem}
\end{minipage}
\end{center}
\end{figure}

The Borexino detector is placed in the underground Laboratori Nazionali del Gran Sasso (LNGS) in Italy. The mountains provide a shielding of 1400 m
of rock ($\sim$ 3800 m.w.e). The detection mechanism for neutrinos is electric scattering on electrons in the liquid scintillator target. The scintillator is
a mixture of pseudocomene (PC, 1,2,4-trimethylbenzene) and PPO (2,5-diphenyloxazole at a concentration of 1.5 g/l), a fluorescent dye.
To reach the necessary low background and trigger rate in the detector, a graded shielding is used for the detector (Fig.~\ref{shem}).
The described scintillator is placed in the innermost volume enclosed by a spherical, 125 $\mu$m thick nylon Inner Vessel (IV) with a radius of 4.25 m. The scintillation light
generated by neutrino events is monitored by 2212 8'' photomultiplier tubes (PMTs), type ETL9351, mounted on a Stainless Steel Sphere (SSS)
with a radius of 6.85 m. Most of the PMTs (1828) are equipped with aluminium light concentrators focusing their field of view on the IV.
To shield the scintillator of radiation emitted by the PMTs and the SSS a buffer liquid ($\sim$ 1000 t of pure PC and 5.0 g/l DMP (dimethylphthalate) used
as quencher) is contained in between them. In addition, an Outer Vessel (OV), radius 5.50m, is included in the system. It prevents the diffiusion of radioactive
$^{222}$Rn, emanated from the PMTs, into the IV. The Inner Detector (ID) is fully enclosed by the outermost shield, formed by $\sim$ 2000 t of ultra pure water.
The cylindrical dome (called Outer Detector (OD)) has a height of 16.9 m and a diameter of 18 m.
208 PMTs (same type as used for the ID, but all without light concentrators) are mounted on the SSS pointing outward and detect the Cherenkov Light generated by
through-going muons. The OD is therefore an active muon veto.\newline
A more detailed description can be found in publications~\cite{ali1,ali2}.

\section{Radiopurity and background levels}\label{sec:purity}
Because it is not possible to distinguish neutrino recoil electrons from electrons due to the natural radioactivity and neutrino rates are low (e.g. $^{7}$Be
neutrinos $\sim$ 50 counts/day/100t), a very high level of radioactive purity has to be achieved for the detector. First of all, the scintillator itself has to be very
clean. A special purification strategy removes the most dangerous contaminants~\cite{ali3}. The contamination due to $^{238}$U and $^{232}$Th ((1.6$\pm$0.1)
$\cdot10^{-17}$g/g and (6.8$\pm$1.3)$\cdot10^{-18}$g/g) is one level of magnitude lower than the foreseen goal.\newline
The contamination of $^{85}$Kr, that features a $\beta$-spectrum overlapping with the $^{7}$Be recoil spectrum, is determined to be $(29\pm14)$ counts/day/100tons.
The rather large uncertainty arises from the small branching ratio of the $\beta\gamma$-coincidence decay used to determin the $^{85}$Kr concentration.\newline
The $\alpha$-emitter $^{210}$Po has to be taken into account as its visible decay energy lies in the neutrino window. In the beginning of data taking its activity was
determined to 8000 counts/day/100tons. The rate decays according to the isotope lifetime of 200 days. This means that the concentration of its parent nuclei
($^{210}$Bi and $^{210}$Pb) in the scintillator is much lower.\newline
A large part of background is formed by $^{14}$C contained in the organic scintillator. The analysis threshold at 200 keV corresponds to its spectral endpoint
at 156 keV.\newline
Another relevant background, located in the energy region from 1-2 MeV, is cosmogenically produced $^{11}$C. Its rate is determined to 25 counts/day/100tons and
is at the upper limit of previous studies~\cite{hag,bac}.
Last, after the filling a radon contamination of a few counts per week is present. The source is emanation from the IV.

\section{Data analysis}\label{sec:data}
The trigger to read out the detector is formed in two different ways. Either 25 PMTs in the ID (corresponding to $\sim$ 50 keV of visible energy) fire in a time window
of 60 ns, or 6 PMTs in the OD in 150ns. In any case both detectors (ID and OD) are read out.\newline
The charge corresponding to the amount of photons collected by all PMTs is given in photo electrons (pe). In the analysis the light yield is determined by a fit to the
observed $^{11}$C spectrum or to the end-point of the $^{14}$C spectrum. A value of 500 pe/MeV of deposited energy is the consistent result of the
different methods.

\subsection{The $^{7}$Be neutrino spectrum}\label{sec:7be}
\begin{figure}[h]
\begin{center}
\begin{minipage}{0.8\textwidth}
\centering{\includegraphics[width=0.75\textwidth]{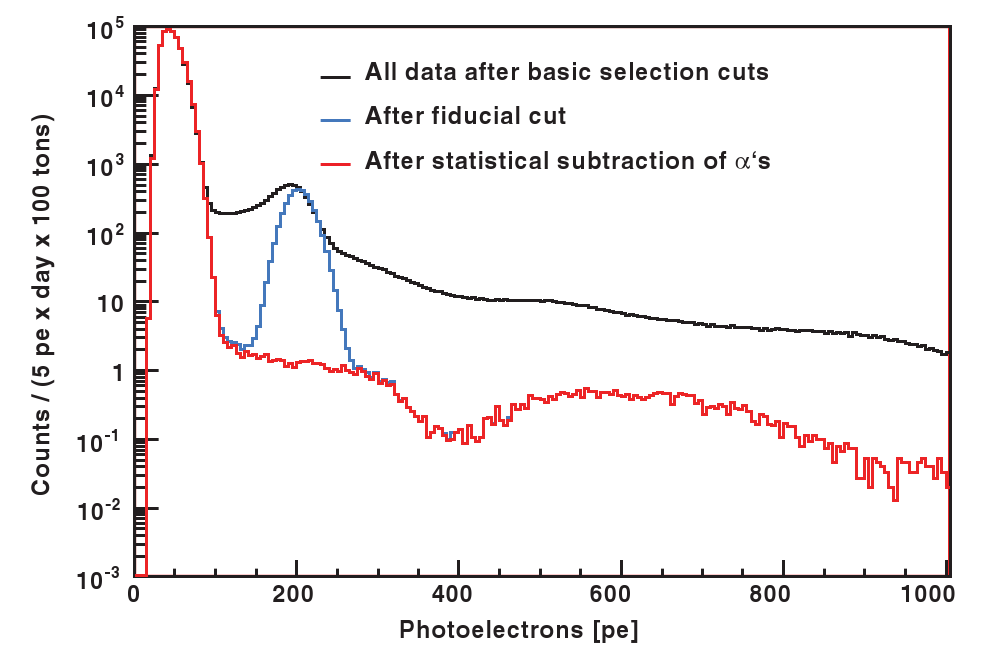}}
\caption{Raw charge spectrum used for determination of $^{7}$Be neutrinos. Different effects of applied analysis cuts.}\label{7beraw}
\end{minipage}
\end{center}
\end{figure}
\begin{figure}[h]
\begin{center}
\begin{minipage}{0.8\textwidth}
\centering{\includegraphics[width=0.75\textwidth]{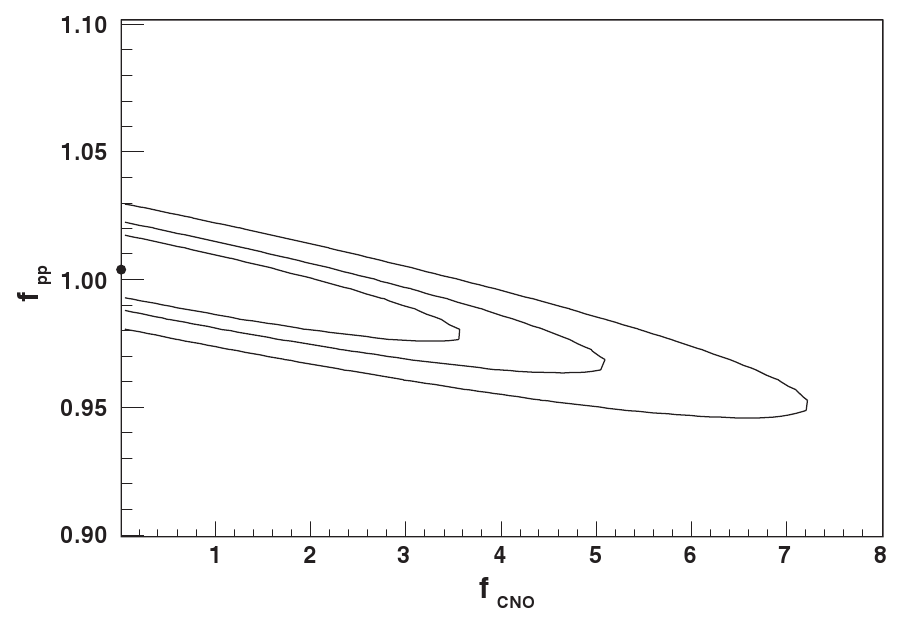}}
\caption{Determination of flux normalization constants for pp
and CNO solar neutrinos without luminosity constraint, $f_{pp}$ and $f_{CNO}$ (68\%, 90\%, and 99\%
C.L.).}\label{ppcno}
\end{minipage}
\end{center}
\end{figure}

The curves presented in Fig.~\ref{7beraw} are the results of 192 live days of measurement. The black line demonstrates the raw data with the 3 most basic cuts
already applied. This means, that only single clustered events are accepted in order to reject pile up or fast coincidences. Second all muons are neglected.
This tagging can be performed using the OD as an active muon veto. The efficiency is $99.5\%$. In addition,
muons and neutrinos can be distinguished by pulse shape analysis using the ID. The remaining inefficiency is reduced to less than $10^{-3}$.
And third, as a logical consequence, all events detected in a 2 ms time window after each muon, are rejected in order to avoid muon induced secondaries. The
blue curve is obtained using a fiducial volume cut to reject external $\gamma$ background: The allowed detector volume corresponds to the innermost 100 tons
($\sim$ 3.25 m radius). Also the radon induced $^{214}$Bi-$^{214}$Po coincidences are removed from data.\newline
The large $^{14}$C peak remains at the lower energy range of the observed spectrum, which therefore will not influence the results of the $^{7}$Be neutrinos.
Another remaining obvious part is the still present $^{210}$Po peak at about 190 pe. It can be rejected by $\alpha/\beta$ pulse shape discrimination (red curve).
In both cases, blue and red line, the compton like edge of the $^{7}$Be neutrinos (300-350 pe) and the $^{11}$C spectrum (400-800 pe) becomes visible.\newline
\begin{figure}[h]
\begin{center}
\begin{minipage}{0.8\textwidth}
\centering{\includegraphics[width=0.75\textwidth]{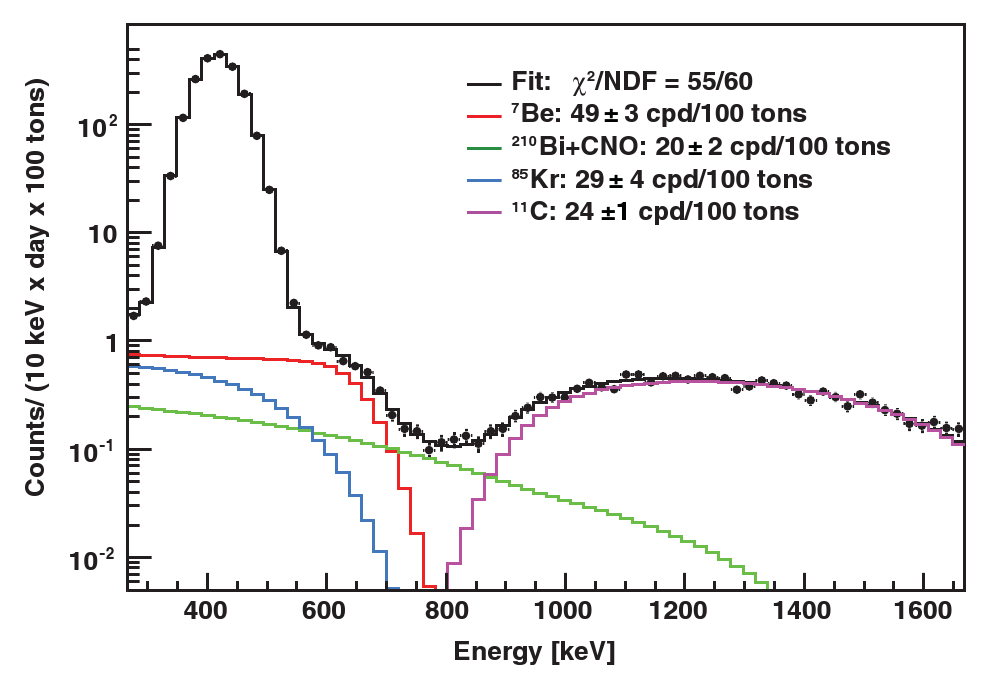}}
\caption{Spectrum and fits of $^{7}$Be neutrinos. $^{210}$Po-peak still present.}\label{7be}
\end{minipage}
\end{center}
\end{figure}
In Fig.~\ref{7be}, as final result, the fits applied to the remaining data are presented. Two independent methods are used for this evaluation and come to consistent
results (expressed in counts/day/100tons, statistical errors only): 49 $\pm$ 3 for $^{7}$Be, 20 $\pm$ 2 for a sum of CNO-neutrinos and $^{210}$Bi,
29 $\pm$ 4 for $^{85}$Kr, 24 $\pm$ 1 for $^{11}$C.\newline
With an estimation of the systematic error (8.5 $\%$) a $^{7}$Be neutrino rate of ($49 \pm 3_{stat} \pm 4_{sys}$) counts/day/100tons can be announced. The
rate expected without neutrino oscillation for the high metallicity~\cite{gre} Standard Solar Model (SSM)~\cite{beh} (BS07(GS98)) is (74 $\pm$ 4) counts/day/100tons.
The flux normalization constant $f$, the ratio between the measured MSW-LMA scenario~\cite{mik,wolf,hol} and the predicted BS07(GS(98)) neutrino flux, is
calculated to $f_{Be}=1.02\pm0.10$. The observed survival probability at the $^{7}$Be energy of 862 keV is $P_{ee}=0.56\pm0.10$. The no oscillation hypothesis
($P_{ee}=1$) is rejected at 4$\sigma$ C.L.. So Borexino confirms the MSW-LMA neutrino oscillation scenario and provides the first direct measurement in the low-energy
vacuum MSW regime~\cite{behc}.\newline
The new Borexino results can be used along with the results of radiochemical experiments~\cite{land,hamp,abd,alt} and SNO~\cite{ahm,aha} to calculate new limits
on the pp- and CNO-neutrino fluxes. The results are at present the best experimental limits and are shown in Fig.~\ref{ppcno} for different C.L..
Including in addition the solar luminosity constraint we determine $f_{pp}=1.005^{+0.008}_{-0.020} (1\sigma)$ and $f_{CNO} < 3.80$ $(90\% C.L.)$, using
the 1-D $\chi^{2}$-profile method~\cite{yao}.
The results of $f_{CNO}$ can be transformed into a contribution to the total solar neutrino luminosity of less than $3.3\%$ $(90\% C.L.)$.

\subsection{The $^{8}$B neutrino spectrum}\label{sec:8B}
After 246 live days of measurement a preliminary analysis of the $^{8}$B neutrino rate is performed. Due to the low threshold of the Borexino detector, it is the first
real-time experiment able to measure the flux down to an energy limit of 2.8 MeV, in comparison to the water-Cherenkov detectors with a threshold of 5MeV.\newline
For the analysis different cuts, similar to the ones used for the $^{7}$Be neutrino analysis, have to be applied. As main background for energies higher 2.8 MeV, muons
and their secondaries have to be identified. Radon emanation from the nylon vessels is reduced using the fiducial volume
cut. Applying a 5 s time and different position cuts short-lived and long-lived cosmogenic isotopes can be rejected. Last the bulk $^{232}$Th
and the buffer $^{208}$Tl contermination have to be excluded.\newline
As result of this preliminary analysis, a $^{8}$B neutrino rate of ($0.26 \pm 0.04_{stat} \pm 0.02_{sys}$) counts per day and 100 tons down to an energy of 2.8 MeV can be
reported. The corresponding $\nu_{e}$ flux of ($2.65\pm0.44_{stat}\pm0.18_{sys})\cdot10^{6}$ cm$^{-2}$s$^{-1}$ is in agreement with the MSW-LMA SSM prediction
and SNO or Super-Kamiokande results~\cite{aha,hos}.
Assuming the BS07(GS98) SSM, the mean survival probability is 0.35$\pm$0.10 at the effective energy of 8.6 MeV. The no-oscillation scenario can be excluded using this
preliminary analysis at a 4.2 $\sigma$ C.L..\newline
Borexino is the first experiment with the ablility to simultaneous measure solar
neutrinos from the vacuum region ($^{7}$Be) and from the matter-enhanced oscillation region ($^{8}$B). The ratio between the two different survival probabilities for
$^{7}$Be and $^{8}$B neutrinos is 1.60$\pm$0.33, 1.8 $\sigma$ different from unity. Still at relatively low significance, this result confirms the transition between low
energy vacuum dominated and high energy matter enhanced solar neutrino oscillation, as predicted by the MSW-LMA solution.

\section{Future Objectives}\label{sec:out}
At present a calibration campain is performed in order to allow a measurement of the $^{7}$Be flux at $5\%$ accuracy by reduction of systematic uncertaincies.
In addition to gather more statistics for the measurement of
$^{7}$Be and $^{8}$B, the attention is turned to the direct, and unprecedented, measurement of the low energetic pep- and CNO-neutrinos. The possibility to measure
the pp-neutrino flux is investigated. Further aims of Borexino are the observation of geo-neutrinos and (anti)neutrinos from a galactic supernova. For instance Borexino will
join the SNEWS (supernova early warning system), foreseeably during 2009.

\end{document}